# A Novel Antimicrobial Electrochemical Glucose Biosensor Based on Silver-Prussian Blue Modified TiO2 Nanotube Arrays

Nasim Farajpour[2], Ram Deivanayagam[3], Abhijit Phakatkar[3], Surya Narayanan[1], Reza Shahbazian-Yassar[3]*, Tolou Shokuhfar[1]*

# A Novel Antimicrobial Electrochemical Glucose Biosensor Based on Silver-Prussian Blue Modified TiO$_2$ Nanotube Arrays

Nasim Farajpour[2], Ram Deivanayagam[3], Abhijit Phakatkar[3], Surya Narayanan[1], Reza Shahbazian-Yassar[3*], Tolou Shokuhfar[1*]

[1]Department of Bioengineering, University of Illinois at Chicago, Chicago, IL 60607, USA

[2]Department of Electrical and Computer Engineering, University of Illinois at Chicago, Chicago, IL60607, USA

[3]Department of Mechanical and Industrial Engineering, University of Illinois at Chicago, Chicago, IL 60607, USA

**\***tolou@uic.edu**;** rsyassar@uic.edu

**Abstract**. Glucose biosensors play an important role in the diagnosis and continued monitoring of the disease, diabetes mellitus. This report proposes the development of a novel enzymatic electrochemical glucose biosensor based on TiO$_2$ nanotubes modified by AgO and Prussian blue (PB) nanoparticles (NPs), which has an additional advantage of possessing antimicrobial properties for implantable biosensor applications. In this study, we developed two high performance glucose biosensors based on the immobilization of glucose oxidase (GOx) onto Prussian blue (PB) modified TiO$_2$ nanotube arrays functionalized by Au and AgO NPs. AgO-deposited TiO$_2$ nanotubes were synthesized through an electrochemical anodization process followed by Ag electroplating process in the same electrolyte. Deposition of PB particles was performed from an acidic ferricyanide solution. The surface morphology and elemental composition of the two fabricated biosensors were investigated by scanning electron microscopy (SEM) and energy-dispersive X-ray spectroscopy (EDS) which indicate the successful deposition of Au and AgO nanoparticles as well as PB nanocrystals. Cyclic voltammetry and chronoamperometry were used to investigate the performance of the modified electrochemical biosensors. The results show that the developed electrochemical biosensors display excellent properties in terms of electron transmission, low detection limit as well as high stability for the determination of glucose. Under the optimized conditions, the amperometric response shows a linear dependence on the glucose concentration to a detection limit down to 4.91 µM with sensitivity of 185.1 mA M$^{-1}$ cm$^{-2}$ in Au modified biosensor and detection limit of 58.7 µM with 29.1 mA M$^{-1}$ cm$^{-2}$ sensitivity in AgO modified biosensor.

1. Introduction

Diabetes mellitus is a chronic disease threatening the health of an increasing number of individuals on a global scale [1]. It is caused by insufficient production or function of the hormone - insulin - produced by the pancreas, leading to a glucose level beyond the standard range of 80-120 mg/DL in blood plasma [2]. Accurate measurement of the blood glucose levels is important for the diagnosis of this chronic condition. There have been numerous attempts in the recent decades to develop sensitive and reliable glucose biosensors, which demonstrate the growing demand for those devices in the community [3]. Recently, the development of biosensors based on metal oxides such as titanium oxide has attracted considerable attention in detecting glucose levels [4].

Titanium oxide nanotubes ($TiO_2$ NTs) are one of the most commonly used nanomaterials in a variety of biosensor applications, due to their distinctive characteristics such as large active surface area, great thermal and chemical stability, and excellent mechanical strength[5][6]. These inorganic nanostructures have been used as a suitable matrix for enzyme immobilization to stabilize the enzyme as well as to retain the functionality of the enzyme [7].

Previous studies have shown that noble metals modified $TiO_2$ nanotubes demonstrate better catalyzation properties due to their beneficial features such as biocompatibility and high conductivity[8]. They can be considered as catalyzers because of their ability to improve electrochemical reaction rate as well as electron transfer channels between the enzyme and active surface of the electrode[8][9]. Gold (Au) and Silver (Ag) nanoparticles are among the most promising metal nanoparticles for biosensing applications due to many favorable attributes such as excellent conductivity, simplicity of fabrication, and cost efficiency[9][10].

The mechanism of glucose measurement is based on the oxidization of β-D-glucose by dissolved molecular oxygen with catalyzing immobilized Glucose oxide (GOx) enzyme which results in the generation of gluconic acid and hydrogen peroxide ($H_2O_2$), as shown in equation (1) [11]

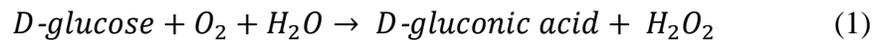
$$D\text{-}glucose + O_2 + H_2O \rightarrow D\text{-}gluconic\ acid + H_2O_2 \qquad (1)$$

During the process of glucose measurement, a high working potential of 0.6 V is required for $H_2O_2$ reduction. This potential also causes electrochemical oxidation of other active species in blood plasma, e.g., uric acid and ascorbic acid which in turn cause an interfering amperometric signal current, thereby decreasing the overall selectivity and accuracy of glucose biosensor [12]. To overcome these issues, peroxidase is typically utilized to cooperate with enzyme glucose oxidase in formation of an oxidase/peroxidase system. Hence, the electrochemical oxidation changes to a reduction process that occurs in a much less potential and significantly improve selectivity and sensitivity of the glucose biosensor[13].

Prussian blue (PB) [$Fe_4[Fe(CN)_6]_3$] is considered an "artificial enzyme peroxide" in electrochemical reduction of hydrogen peroxide due to its high selectivity and electrocatalytic function specific to $H_2O_2$ reduction [14] [15] . Prussian blue has been investigated as a promising precursor and template to be used in electrocatalysis owing to its easy preparation and low cost [16]. The traditional Prussian blue synthetic method is based on mixture of ferric ($Fe^{3+}$) and ferricyanide ($[Fe(CN)_6]^{3-}$) ions aqueous solution[18]. Moreover, findings indicate the significant effects of Au and Ag nanoparticles to improve PB reaction rate and growth process[19][20].

In this article, we have developed two high performance glucose biosensors based on immobilization of enzyme glucose oxide onto Au and Ag oxide (AgO) modified titanium oxide nanotubes ($TiO_2$ NTs), followed by the deposition of Prussian blue (PB) as the electron transfer mediator. The surface morphology and elemental composition of the two fabricated biosensors were investigated by scanning electron microscopy (SEM) and energy-dispersive X-ray spectroscopy (EDS) techniques which confirmed the successful deposition of Au and AgO nanoparticles as well as PB nanocrystals. The electrochemical measurement of the developed biosensors demonstrates high sensitivity and selectivity, excellent biological stability, and low detection limit towards determination of glucose concentration.

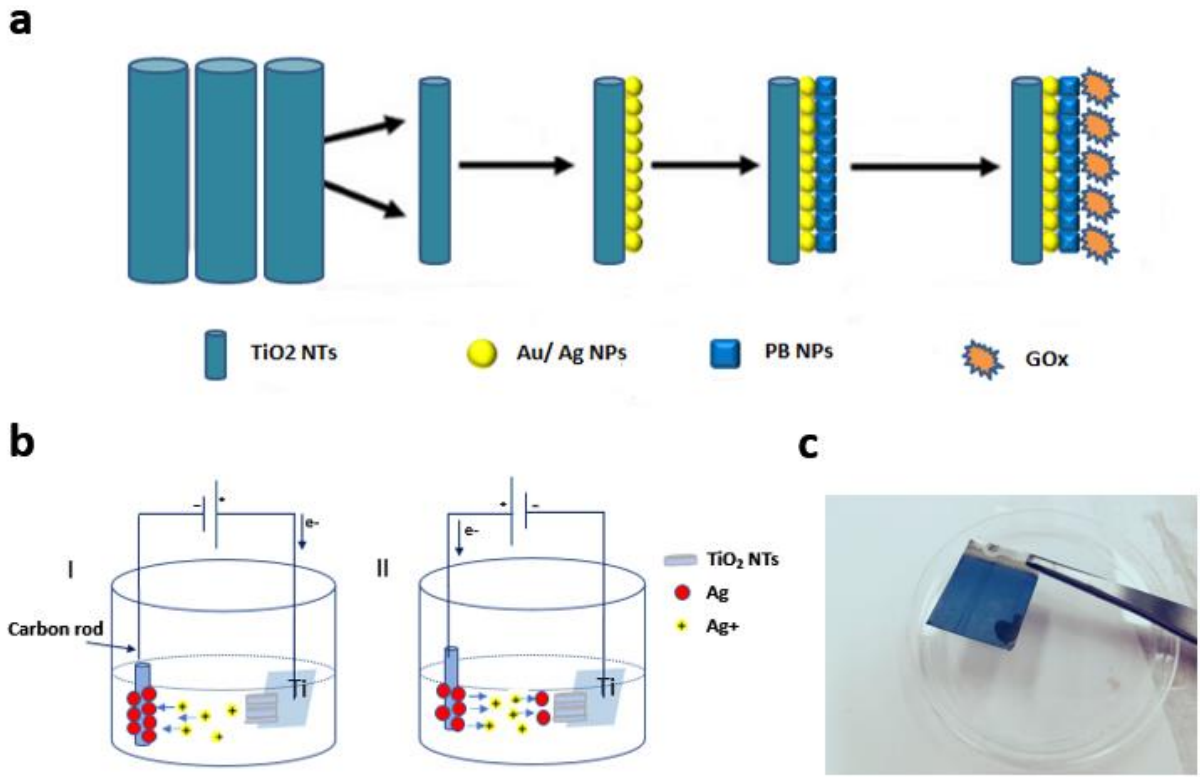

**Figure 1** (a) General Schematic for preparing bioenzyme electrode based on TiO$_2$ NTs. (b) Schematic diagram of AgO-deposited TiO$_2$ NTs formation. (b) (I) Anodization process is shown. Ag ions moved toward carbon cathode, which was negatively biased with respect to Ti. Ag atoms were deposited on the cathode while TiO$_2$ NTs were formed on Ti foil which was positively biased. (II) Ag electroplating process is shown. The polarity was reversed and Ag$^+$ ions moved toward negatively biased Ti with respect to carbon cathode. Ag atoms deposited on the TiO$_2$ NTs layer on Ti. (c) Dark blue PB thin film deposition overnight onto the modified electrode surface

## 2. Experimental

2.1. Chemicals and Instruments

Titanium sheets (0.25 mm thickness, 99.7 % purity), Potassium Ferricyanide, Potassium Chloride, 1,2-diaminobenzen (1,2-DAB) and Glucose oxidase (GOx) were purchased from Sigma Aldrich. Ferric (III) Chloride, Silver Flouride, b-D (+) glucose were purchased from Fisher Scientific. All the chemicals were of analytical grade and were used without further purification. All the solutions were prepared with deionized water.

Anodization of $TiO_2$ NTs was performed using traditional two electrode setup with carbon rod as the cathode and Ti foil as the anode in a self-made electrochemical cell (KEITHLEY 2400). The surface morphology of both the electrode samples was characterized through Field Emission Scanning Electron Microscope (JEOL JSM-6320F) and Scanning Electron Microscope (SEM, Raith 100). Elemental compositions of the fabricated electrodes were investigated by energy-dispersive X-ray spectroscopy (EDS) (Hitachi S-3000N VPSEM). Cyclic voltammetry and chronoamperometry measurements were performed with an electrochemical workstation (Bio-Logic VMP3 equipped with EC-Lab software) using a three-electrode setup comprising an Ag/AgCl reference electrode, a Pt foil auxiliary electrode, and either of the two modified Titanium nanotubes electrodes (GOx/PB/Au/$TiO_2$ NTs) and (GOx/PB/AgO/$TiO_2$ NTs) as the working electrodes.

2.2. Fabrication of Titanium NTs

Electrochemical anodization has been used for TiO2 NTs formation. Prior to anodization, pure Ti foil was sonicated in ethanol for 20 minutes followed by rinsing with DI water. The anodization station was a self-made two-electrode electrochemical cell. The Ti foil was used as the anode while a carbon rod electrode served as the cathode. The anodization process was performed at 60 V DC for 2 hours in a freshly prepared electrolyte containing 0.2 wt% $NH_4F$, 10 vol % deionized water and ethylene glycol to form nanotubes on the Titanium substrate surface.

After TiO2 NTs formation, the prepared sample was sonicated in isopropanol and deionized water each for 2 minutes respectively.

In order to synthesize the AgO/TiO$_2$ NTs, electrochemical anodization was used for TiO$_2$ NTs formation and Ag/TiO$_2$ NTs electrode was prepared by following the method in a previous work[21]. Preparation of titanium foil and anodization station was similar to what was described in section 2.2.1. The anodization process was performed at 60 VDC for 90 minutes in a freshly prepared electrolyte containing 0.686 g NH$_4$F, 98 mL ethylene glycol and 2 mL deionized water to form nanotubes on the Titanium substrate surface.

Following formation of TiO$_2$ NTs, Ag$^+$ ions dispersed in the electrolyte solution get adsorbed onto the carbon rod cathode, with the electric field supplying the electrons for Ag deposition. Deposition of Ag nanoparticles was carried out by applying potential of 20 V for 60 seconds together with switching the electrodes. During this procedure, Ag on the cathode loses electrons and gets dissolved into the electrolyte. Under the electric field, Ag$^+$ in the electrolyte migrated into the TiO$_2$ NTs, took up electrons, and formed Ag nanoparticles. After deposition of Ag nanoparticles, the prepared sample was rinsed in isopropanol and deionized water each for 2 minutes.

2.3. Preparation of Prussian blue/Au modified TiO$_2$ NTs

To fabricate Au modified electrode, TiO$_2$ NTs were immersed in a 0.05% PolyDADMAC (PDDA) solution for 3 hours at room temperature. After rinsing in deionized water, the modified electrodes were incubated into 20 mL of the AuNPs colloidal solution (10nm) at 4 °C overnight. For decoration of the AuNPs modified TiO$_2$ NTs electrode with PB NPs, the prepared sample was immersed in an acidic solution containing 0.1 mM K$_3$[Fe(CN)$_6$] + 0.1mM FeCl$_3$ in the presence of 0.1M KCl overnight. Hydrochloric acid was used for adjusting the pH of aqueous solution to 1.6. Deposition of PB nanoparticles was observed by forming a very thin dark blue film on the electrode surface. The same procedure has been used for decoration of the AgO modified TiO$_2$ NTs electrode with PB nanoparticles.

2.4. GOx enzyme immobilization and fabrication of glucose biosensor

For enzyme GOx immobilization, the electrochemical polymerization method was used to obtain GOx/PB/Au/TiO$_2$ NTAs sample electrode. The cyclic voltammetry was conducted in a 0.01 M PBS solution electrolyte containing 30 mg ml of GOx and 10 mM o-Phenylenediamine (PH 7) at sweep voltage of 50 mV s$^{-1}$ for 5-10 cycles. After rinsing with deionized water, the enzyme modified electrode was stocked at 4 ºC.

For AgO modified biosensor preparation, physical adsorption method was used for immobilization of enzyme GOx to obtain GOx/Ag/ TiO$_2$ NTs electrode. Fresh enzyme solution was prepared in a 0.01 M PBS electrolyte containing 20 mg ml of GOx and 10 mM o-Phenylenediamine. Afterward, 50 µL of the solution was dropped onto the PB/Ag/TiO$_2$ NTs electrode and dried in air following which it was kept at 4 ºC for 10 hours. After rinsing with PBS, the enzyme modified electrode was stocked at 4 ºC when not in use.

3. **Results and Discussions**

To accelerate the electron transfer between the electrode and the enzymes in TiO$_2$-based biosensor systems, the TiO$_2$ NTs can be modified by decorating them with metal, metal oxide and other different active catalysts with coordination polymers such as Prussian blue acting as a mediator. In this case, TiO$_2$ NTs, being negatively charged in a neutral solution, get wrapped by cationic polyelectrolyte PDDA via electrostatic interaction. After the formation of positively charged PDDA-modified TiO$_2$ NTs, the negatively charged AuNPs attach themselves to the tube surfaces and walls through the PDDA bridges. The attached AuNPs largely improve the electronic conductivity as well as the photocatalytical activity of the TiO$_2$ NTs.

For AgO modified TiO$_2$ NTs synthesize, surface electroplating method has been used. Electroplating or electrochemical deposition is a process in which metal ions in electrolyte deposit on the electrode surface upon application of an electric field or current. The parameters that control the extent of electroplating are current/voltage source, temperature, plating duration, electrolyte concentration, the distance between electrodes and stirring. As it is depicted in Figure

1b, following formation of TiO$_2$ NTs, Ag$^+$ ions dispersed in the electrolyte solution gets adsorbed onto the carbon rod cathode as the electron was supplied for Ag deposition by the electric field. During Ag deposition procedure, Ag on the cathode loses electrons and gets dissolved into the electrolyte. Under the electric field, Ag$^+$ in the electrolyte migrates into the TiO$_2$ NTs, takes up electrons, and forms Ag nanoparticles.

For PB deposition on conducting substrate, the Au modified and AgO modified TiO$_2$ NTs electrodes were incubated in an acidic ferricyanide solution overnight which resulted in formation of thin PB layer in dark blue color, as shown in Figure 1c. Leakage of PB into solution phase will occur after a few scans at neutral pH due to the relatively weak combination between PB and bare electrode. To suppress the dissolution of PB and thereby improve the stability of PB modified nanotubular electrode, a thin film of o-Phenylenediamine as a nonconducting polymeric layer was deposited on the top of the PB deposited electrodes.

The surface morphology of electrodeposited metal on surface of the nanotubes was determined by Scanning Electron Microscopy (FE-SEM, Hitachi S-4700). Samples were installed on the holder and were partially covered by conductive tape and placed into the chamber with 10KV voltage and 8mm working distance. Since the accumulation of silver ions on the tubes makes its surface uneven, the SEM images did not show the deposition as clearly as in the case of pure nanotubes. To analyze the elemental composition and to make sure that the deposited materials are silver nanoparticles, energy-dispersive X-ray spectroscopy (EDX) was used. SEM images in Figure 2a and 2c show that diffusion of gold, silver and PB nanoparticles on TiO$_2$ NTs surface. The nanoparticles deposition on the TiO$_2$ layer may also be confirmed by the EDX analysis illustrated in Figure 2b and 2d. Au NPs, Ag NPs and PB nanocrystals can be observed on the tube entrance and walls. The results of energy spectrum show the presence of C and Fe elements of PB.

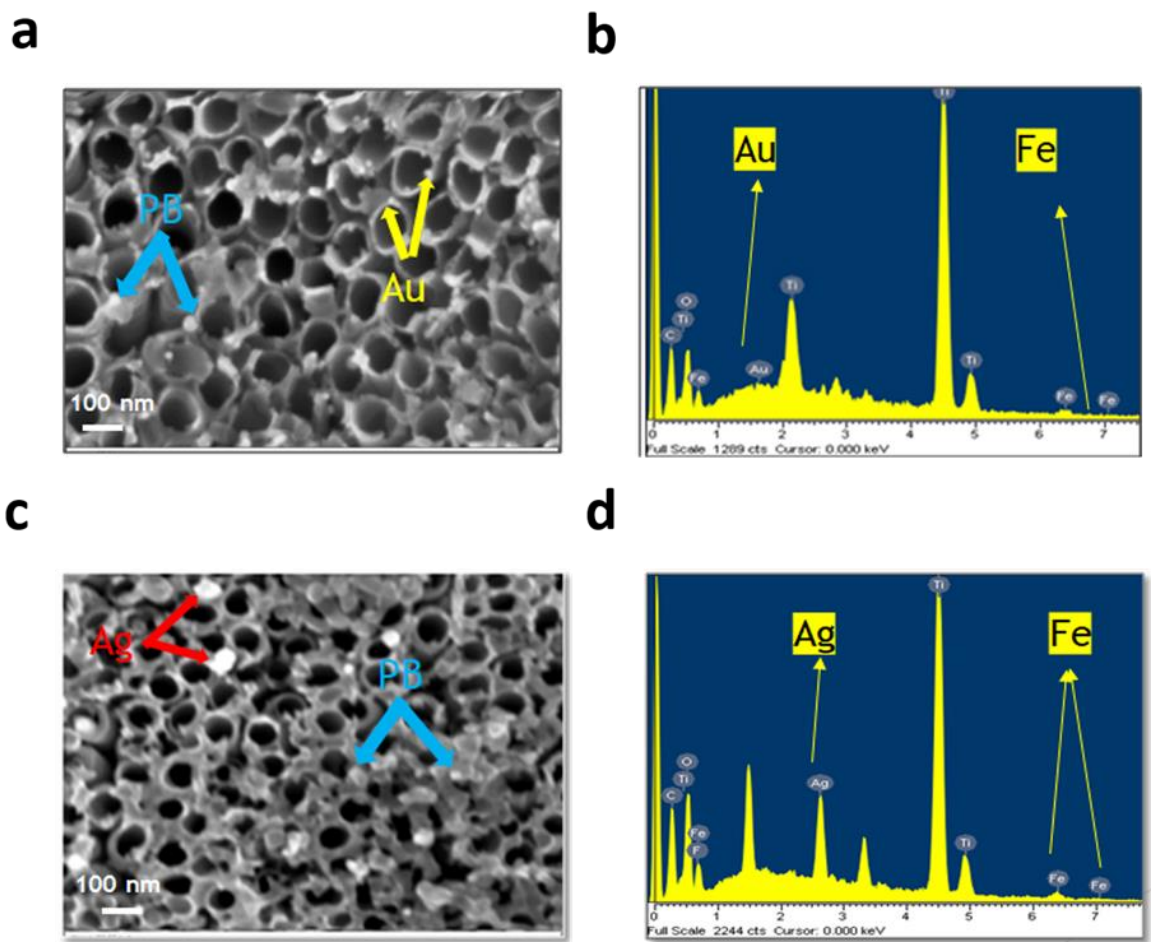

Figure 2 (a) SEM images of Au and PB nanoparticles deposited on to the TiO$_2$ NTs. (b) EDS elemental analysis. (c) SEM Images of Ag and PB nanoparticles deposited on to the TiO$_2$ NTs. (d) EDS elemental analysis

Electrochemical polymerization of Phenylenediamine on PB-Au Modified TiO$_2$ NTs Electrode has shown in Fig. 3a. Due to the presence of nonconducting polymer on electrode surface, the electron transfer between the electrode and the electrolyte is hindered. Consequently, the anodic currents decrease accordingly with the polymerization of o-Phenylenediamine during subsequent CV scans. This tendency has a certain affinity with the previous reports[22][23]. Fig. 3b represents the progress on electropolymerization process to prepare GOx/PB/Au NPs/TiO$_2$ NTs electrode by sweep rate of 50 mV s$^{-1}$ for 1–10 cycles.

The electrochemical activity of the modified TiO$_2$ NTs was investigated by CV as shown in Figure 4a. The PB/AuNP/TiO$_2$ NTs demonstrate a higher cathodic and anodic current response in compare to Au modified TiO$_2$ NTs and simple TiO$_2$ NTs. Pure Ti electrode has been examined as well which exhibit very negligible electrochemical response activity in PBS solution. This confirms the effect of Au and PB NPs in providing better electron transfer and improved redox reaction response of the electrode.

The electrochemical activity and the stability of the fabricated electrode was investigated using CV. As evident in Fig. 4b, the GOx/PB/AuNP/TiO$_2$ NTs electrode exhibits cathodic and anodic currents in 0.01 M PBS solution (pH 6.0, containing 0.1 M KCl) at scan rate of 50 mV s$^{-1}$. Upon addition of 10 mM glucose into the PBS solution, the rapid increases in the cathodic currents appear as a result of the reduction of hydrogen peroxide (H$_2$O$_2$) produced from the enzymatic reaction. The couple of reversible redox peaks located at -0.68 V originated from the transformation process between Prussian blue (PB) and Prussian white (PW). The GOx/PB/AuNP/TiO$_2$ NTs based electrode exhibits a higher response current to 10 mM glucose (curve c) than the one with no glucose (curve b), suggesting that the GOx/Au/PB/TiO$_2$ NTs electrode has a good biocatalytical activity for glucose.

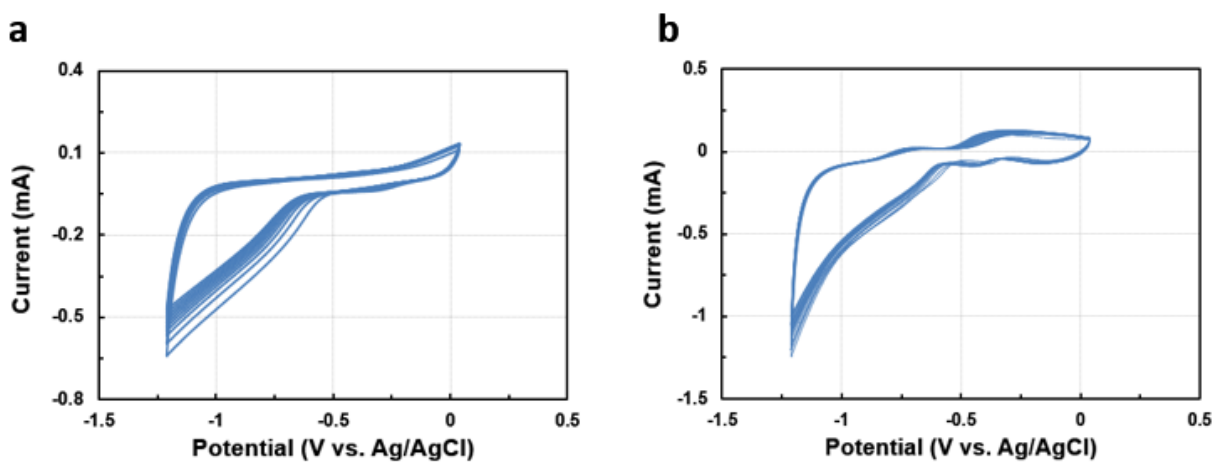

Figure 3 (a) Cyclic voltammograms of Phenylenediamine electrodeposition on PB/Au NPs/TiO$_2$ NTs electrode (b) Cyclic voltammograms of electrodeposition of GOx onto the PB/AuNP/TiO$_2$ NTs electrode via the electropolymerization of Phenylenediamine. Numerals indicated the numbers of electro polymerization cycles. Scan rate: 50 mV s-1.

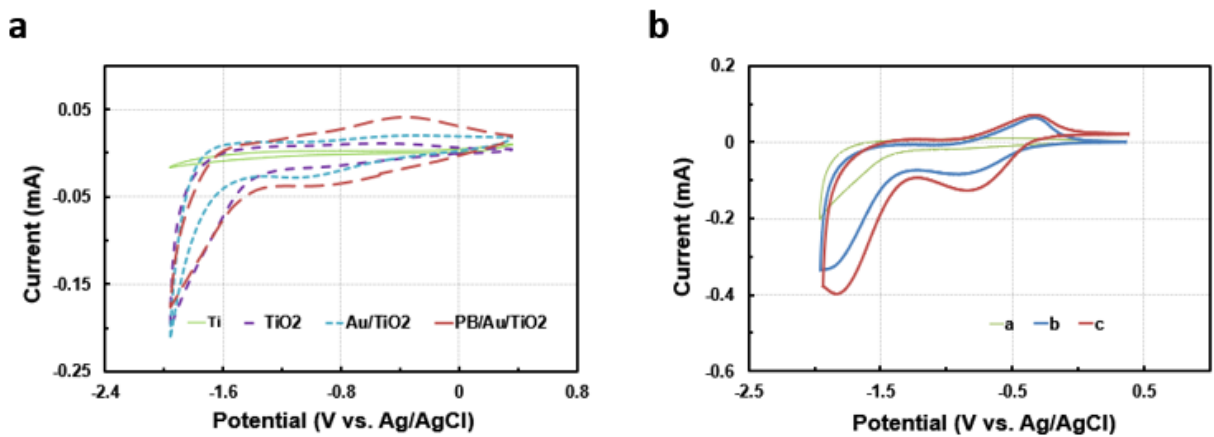

Figure 4 (a) Cyclic voltammograms of the Ti electrode (green plot) in compare with TiO2 NTs electrode (purple plot), Au/TiO2 NTs electrode (blue plot) and PB/Au/TiO2 NTs electrode (red plot) in PBS (pH 6.0). (b) Cyclic voltammograms of the TiO2 electrode (curve a) in compare with PB/Au NPs/TiO2NTs electrode in absence (curve b) and presence (curve c) of 10 mM glucose in PBS (pH 6.0).

In Figure. 5a, the cyclic voltammetry of GOx/PB/Ag NPs/TiO$_2$ NTs electrode at scan rate of 50 mV s$^{-1}$ is shown. The electrode exhibits cathodic and anodic currents in 0.01 M PBS solution (pH 6.0, containing 0.1 M KCl). Upon addition of 10 mM glucose into the PBS solution, rapid increases in the cathodic currents appear due to the reduction of hydrogen peroxide ($H_2O_2$) produced from the enzymatic reaction. The couple of reversible redox peaks located at -0.29 V originated from the transformation process between Prussian blue (PB) and Prussian white (PW). The GOx/PB/Ag NPs/TiO$_2$ NTs based electrode exhibits a higher response current to 10 mM glucose (curve c) than the one with no glucose (curve b), suggesting that the GOx/Ag NPs/PB//TiO$_2$ NTs electrode has a good biocatalytical activity for glucose. Figure 5b shows the cyclic voltammogram of the GOx/PB/Ag NPs/TiO$_2$ NTs modified electrode at scan rate 50 mV s$^{-1}$ in presence of 10 mM glucose in PBS within the voltage range of -0.956 to 0.644 (measured solution pH 6.0).

Figure 6 shows the comparison of the FTIR spectra of TiO$_2$ NT electrode (Curve A) in different bands. The first peak is observed at 3218 cm-1, which corresponds to the hydroxyl group O-H stretching vibration and confirm the adsorption of water molecules in TNTs. The second band is observed around 1610 cm-1 and is related to water Ti-OH bending modes [24][25]. Curve B and

C show the spectra of GOx/PB/AgO/TiO$_2$ NTs and GOx/PB/AuNP/TiO$_2$ NTs respectively. The absorption bands observed at 2082 and 2356 cm-1 are characteristic of PB, and are related to the stretching vibration of the CN group. With addition of GO, the sample exhibits IR absorption peaks at 1506, 1558 and 1652 cm−1 which corresponds to the C–O, C–OH and COO–stretching of GO [26].

The effect of glucose concentration on the current was investigated by adding different concentrations of glucose to the PBS electrolyte. Since the electrochemical signal results from the reduction of PB in the sample, the differences in cathodic current are observed in amperometric measurements. Figure 7a-b shows the chronoamperometric response of the as-prepared electrodes on the successive addition of glucose to the PBS under stirring at an applied potential of -0.25 V. A linear increase in reduction current with the increase in glucose was observed in the range of 0.1 - 0.4 mM.

For the GOx/PB/Au NP/TiO$_2$ NTs based electrode (Fig. 7c), a linear increase in reduction current with the increase in glucose concentration in the range of 0.1– 0.4 mM with a detection limit of 4.91 µM (estimated from 3 times the standard deviation of the blank) was observed. The regression equation of the linear part of the curve was y = -0.1851x - 0.0552, where y represents the current in mA and x the glucose concentration in mM; the R2 was 0.986. The sensitivity was 185.1 mA. For the GOx/PB/AgO NP/TiO$_2$ NTs based electrode (Fig. 7d), the resulted electrode exhibits an acceptable sensing ability. The linear range of such biosensor for the determination of glucose was found with a detection limit of 58.7 µM. The linear regression equation was: y = -0.0291x - 0.0211, where y represents the current in mA and x the glucose concentration in mM; the R2 was 0.981. The sensitivity of this biosensor is 29.1 mA.

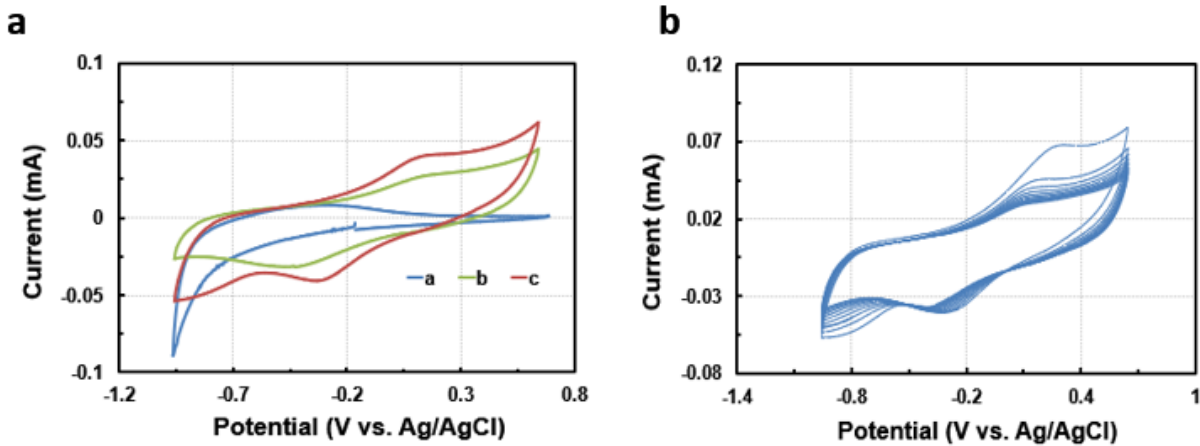

Figure 5 (a) GOx-TiO$_2$ NTs based electrode (curve a), GOx/PB/AgO NP/TiO$_2$ NTs based electrode in absence (curve b) of 10 mM glucose in PBS (pH6.0), GOx/PB/AgO NP/TiO$_2$ NTs based electrode in presence (curve c) of 10 mM glucose in PBS (pH 6.0). (b) Cyclic voltammetric behavior of the GOx/PB/AgO NPs/TiO$_2$ NTs modified electrode in presence of 10 mM glucose in PBS

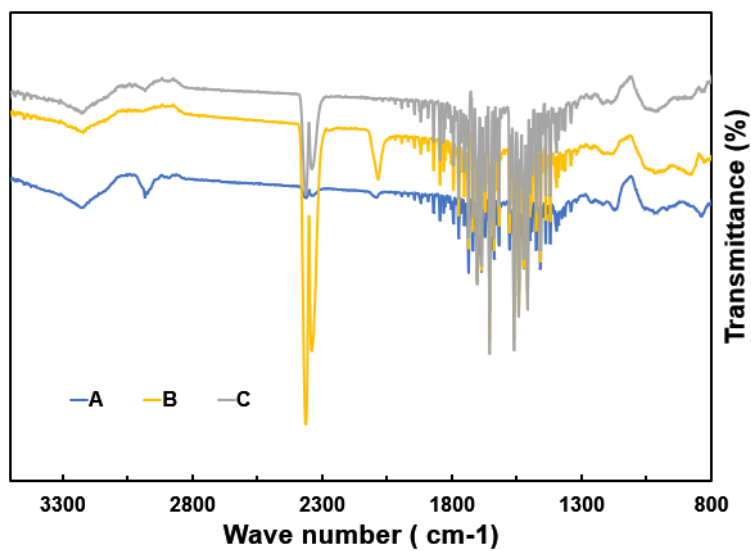

Figure 6 FT-IR spectra of TiO$_2$ NTs (curve A), GOx/PB/AgO NPs/TiO$_2$ NTs (curve B) and GOx/PB/Au NPs/TiO$_2$ NTs sample (curve C)

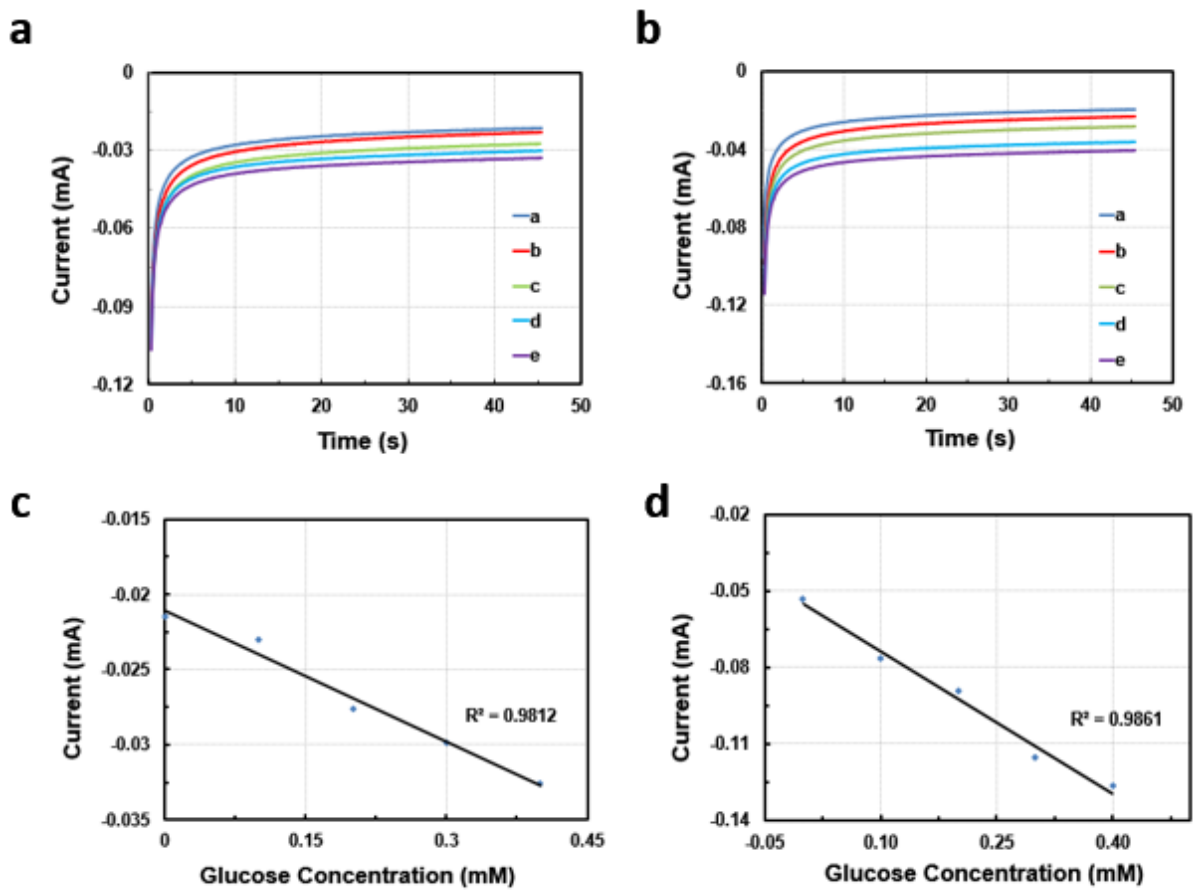

Figure 7 (a) GOx/PB/Au NPs/TiO$_2$ NTs modified electrode (a) in the absence and (b–e) successive addition of glucose at -0.25 V. (b) GOx/PB/AgO NPs/TiO$_2$ NTs modified electrode (a) in the absence and (b–e) successive addition of glucose at -0.25 V. (c) GOx/PB/Au NPs/TiO$_2$ NTs modified electrode (a) in the absence and (b–e) successive addition of glucose at -0.25 V. (d) GOx/PB/AgO NPs/TiO$_2$ NTs modified electrode (a) in the absence and (b–e) successive addition of glucose at -0.25 V

4. **Conclusion and Future work**

In summary, two high performance glucose biosensors based on immobilization of enzyme glucose oxide onto Au and Ag oxide (AgO) modified titanium nanotubes (TiO$_2$ NTs) followed by the deposition of Prussian Blue (PB) as the electron transfer mediator, were developed. The GOx/PB/Au/TiO$_2$ NTs electrode considerably enhanced the current response signal in glucose solution in comparison to a solution with no glucose. Similarly, the GOx/PB/AgO/TiO$_2$ NTs electrode noticeably improved the current response signal in glucose solution. In addition, both electrodes exhibit a significantly increased the current response signal in compared to the GOx/TiO$_2$ NTs electrode. Moreover, the low concentration of hydrogen peroxide demonstrates the absence of any interaction with Ag nanoparticles, which is highly desirable in glucose biosensor applications.

The surface morphology and elemental composition of the two fabricated biosensors show successful deposition of Au and AgO nanoparticles as well as PB nanocrystals. The results show that the developed electrochemical biosensors display good stability, low detection limit as well as high reproducibility for the determination of glucose. Under optimized conditions, the reported biosensors exhibit a good linear response towards glucose concentration in the range of 0.1 to 0.4 mM with a detection limit down to 4.91 µM and 58.7 µM, sensitivity of 185.1 mA M$^{-1}$ cm$^{-2}$ and 29.1 mA M$^{-1}$ cm$^{-2}$ for Au NPs and AgO NPs modified biosensors respectively.

In conclusion, we report a novel method for developing an enzymatic electrochemical glucose biosensor based on assembling of AgO NPs and PB nanocrystals on Titanium NTs. the biocompatibility of Ag NPs and excellent electrocatalytic properties of PB coupled with the large surface area of nanotubular TiO$_2$ NTs structure, lead to enhanced analytical performance of glucose biosensor in terms of high sensitivity, reliable selectivity, quick response time, and high stability.

# References

[1] C. C. Cowie *et al.*, "Prevalence of diabetes and high risk for diabetes using A1C criteria in the U.S. population in 1988-2006.," *Diabetes Care*, vol. 33, no. 3, pp. 562–8, Mar. 2010.

[2] J. Wang*, "Electrochemical Glucose Biosensors," 2007.

[3] M. M. Rahman, A. J. S. Ahammad, J.-H. Jin, S. J. Ahn, and J.-J. Lee, "A Comprehensive Review of Glucose Biosensors Based on Nanostructured Metal-Oxides," *Sensors*, vol. 10, no. 5, pp. 4855–4886, 2010.

[4] S. N. A. Mohd Yazid, I. Md Isa, S. Abu Bakar, N. Hashim, and S. Ab Ghani, "A Review of Glucose Biosensors Based on Graphene/Metal Oxide Nanomaterials," *Anal. Lett.*, vol. 47, no. 11, pp. 1821–1834, 2014.

[5] M. Artigues, J. Abellà, and S. Colominas, "Analytical Parameters of an Amperometric Glucose Biosensor for Fast Analysis in Food Samples," *Sensors*, vol. 17, no. 11, p. 2620, Nov. 2017.

[6] P. Benvenuto, A. K. M. Kafi, and A. Chen, "High performance glucose biosensor based on the immobilization of glucose oxidase onto modified titania nanotube arrays," *J. Electroanal. Chem.*, vol. 627, no. 1–2, pp. 76–81, Mar. 2009.

[7] M. N. Gupta, M. Kaloti, M. Kapoor, and K. Solanki, "Nanomaterials as Matrices for Enzyme Immobilization," *Artif. Cells, Blood Substitutes, Biotechnol.*, vol. 39, no. 2, pp. 98–109, Apr. 2011.

[8] P. Norouzi, F. Faridbod, B. Larijani, and M. R. Ganjali, "Glucose biosensor based on MWCNTs-gold nanoparticles in a nafion film on the glassy carbon electrode using flow injection FFT continuous cyclic voltammetry," *Int. J. Electrochem. Sci.*, vol. 5, no. 9, pp. 1213–1224, 2010.

[9] N. German, J. Voronovic, A. Ramanavicius, and A. Ramanavicienea, "Gold nanoparticles and polypyrrole for glucose biosensor design," *Procedia Eng.*, vol. 47, no. 5, pp. 482–485, 2012.

[10] X. Gan, T. Liu, X. Zhu, and G. Li, "An electrochemical biosensor for nitric oxide based on silver nanoparticles and hemoglobin.," *Anal. Sci.*, vol. 20, no. September, pp. 1271–1275, 2004.

[11] M. M. Rahman, A. J. S. Ahammad, J.-H. Jin, S. J. Ahn, and J.-J. Lee, "A comprehensive review of glucose biosensors based on nanostructured metal-oxides.," *Sensors (Basel).*, vol. 10, no. 5, pp. 4855–86, 2010.

[12] K. Tian, M. Prestgard, and A. Tiwari, "A review of recent advances in nonenzymatic glucose sensors," *Mater. Sci. Eng. C*, vol. 41, pp. 100–118, Aug. 2014.

[13] F. Ricci and G. Palleschi, "Sensor and biosensor preparation, optimisation and applications of Prussian Blue modified electrodes," *Biosens. Bioelectron.*, vol. 21, no. 3, pp. 389–407, Sep. 2005.

[14] A. A. Karyakin, O. V. Gitelmacher, and E. E. Karyakina, "Prussian Blue-Based First-Generation Biosensor. A Sensitive Amperometric Electrode for Glucose," *Anal. Chem.*, vol. 67, no. 14, pp. 2419–2423, Jul. 1995.

[15] V. D. Neff, "Electrochemical Oxidation and Reduction of Thin Films of Prussian Blue," *J. Electrochem. Soc.*, vol. 125, no. 6, p. 886, Jun. 1978.

[16] C. Deng and D. Wang, "Functional Electrocatalysts Derived from Prussian Blue and its Analogues for Metal-Air Batteries: Progress and Prospects," *Batter. Supercaps*, vol. 2, no. 4, pp. 290–310, Apr. 2019.

[17] Y. Yao, X. Bai, and K.-K. Shiu, "Spontaneous Deposition of Prussian Blue on Multi-

Walled Carbon Nanotubes and the Application in an Amperometric Biosensor," *Nanomaterials*, vol. 2, no. 4, pp. 428–444, Nov. 2012.

[18] A. A. Karyakin, "Prussian Blue and Its Analogues: Electrochemistry and Analytical Applications," *Electroanalysis*, vol. 13, no. 10, pp. 813–819, Jun. 2001.

[19] J.-D. Qiu, H.-Z. Peng, R.-P. Liang, J. Li, and X.-H. Xia, "Synthesis, Characterization, and Immobilization of Prussian Blue-Modified Au Nanoparticles: Application to Electrocatalytic Reduction of $H_2O_2$," *Langmuir*, vol. 23, no. 4, pp. 2133–2137, Feb. 2007.

[20] J. Zhao *et al.*, "Amperometric detection of hydrazine utilizing synergistic action of prussian blue @ silver nanoparticles / graphite felt modified electrode," *Electrochim. Acta*, vol. 171, pp. 121–127, Jul. 2015.

[21] Y. Zhao *et al.*, "Facile electrochemical synthesis of antimicrobial $TiO_2$ nanotube arrays," *Int. J. Nanomedicine*, vol. 9, pp. 5177–5187, 2014.

[22] T. Homma, D. Sumita, M. Kondo, T. Kuwahara, and M. Shimomura, "Amperometric glucose sensing with polyaniline/poly(acrylic acid) composite film bearing covalently-immobilized glucose oxidase: A novel method combining enzymatic glucose oxidation and cathodic O2 reduction," *J. Electroanal. Chem.*, vol. 712, pp. 119–123, Jan. 2014.

[23] Z.-D. Gao, Y. Qu, T. Li, N. K. Shrestha, and Y.-Y. Song, "Development of amperometric glucose biosensor based on Prussian Blue functionlized TiO2 nanotube arrays.," *Sci. Rep.*, vol. 4, p. 6891, Nov. 2014.

[24] ‡ Nadica D. Abazović, ‡ Mirjana I. Čomor, ‡ Miroslav D. Dramićanin, ‡ Dragana J. Jovanović, § and S. Phillip Ahrenkiel, and ‡ Jovan M. Nedeljković*, "Photoluminescence of Anatase and Rutile TiO2 Particles†," 2006.

[25] S. Mugundan, B. Rajamannan, G. Viruthagiri, N. Shanmugam, R. Gobi, and P. Praveen, "Synthesis and characterization of undoped and cobalt-doped TiO2 nanoparticles via sol–gel technique," *Appl. Nanosci.*, vol. 5, no. 4, pp. 449–456, Apr. 2015.

[26] Y. Zhang and C. Pan, "TiO2/graphene composite from thermal reaction of graphene oxide and its photocatalytic activity in visible light," *J. Mater. Sci.*, vol. 46, no. 8, pp. 2622–2626, Apr. 2011.